\documentclass[letterpaper,12pt]{article}

\pdfoutput=1
\usepackage{amsmath}
\usepackage{hyperref}

\makeatletter
\renewcommand\section{\@startsection {section}{1}{\z@}%
{-3.5ex \@plus -1ex \@minus -0.2ex}%
{2.3ex \@plus 0.2ex}%
{\normalfont\normalsize\bfseries}}

\renewcommand\subsection{\@startsection{subsection}{2}{\z@}%
{-3.25ex \@plus -1ex \@minus -0.2ex}%
{1.5ex \@plus 0.2ex}%
{\normalfont\normalsize\bfseries}}

\def\@seccntformat#1{\csname the#1\endcsname.\quad}
\makeatother

\newcommand{\oline}[1]{\overline{\mkern-1.0mu#1\mkern0.0mu}}
\newcommand{\xoline}[1]{\hspace{0.30em}\overline{\mkern-5.0mu#1\mkern-2.0mu}\hspace{0.12em}}
\newcommand{\uline}[1]{\underline{\mkern0.0mu#1\mkern-1.0mu}}
\newcommand\redots{\makebox[0.85em][c]{.\hfil.\hfil.}}

\begin{document}

\setlength{\baselineskip}{4.5ex}

\noindent
{\LARGE \bf On a generalised form of subjective}\\[2ex]
{\LARGE \bf probability}\\[3ex]

\noindent
{\bf Russell J. Bowater}\\
\emph{Independent researcher, Sartre 47, Acatlima, Huajuapan de Le\'{o}n, Oaxaca, C.P.\ 69004,
Mexico. Email address: as given on arXiv.org. Twitter profile:
\href{https://twitter.com/naked_statist}{@naked\_statist}\\ Personal website:
\href{https://sites.google.com/site/bowaterfospage}{sites.google.com/site/bowaterfospage}}
\\[2ex]

\noindent
{\small \bf Abstract:}
{\small
This paper is motivated by the questions of how to give the concept of probability an adequate
real-world meaning, and how to explain a certain type of phenomenon that can be found, for
instance, in Ellsberg's paradox. It attempts to answer these questions by constructing an
alternative theory to one that was proposed in earlier papers on the basis of various important
criticisms that were raised against this earlier theory. The conceptual principles of the
corresponding definition of probability are laid out and explained in detail. In particular, what
is required to fully specify a probability distribution under this definition is not just the
distribution function of the variable concerned, but also an assessment of the internal and/or the
external strength of this function relative to other distribution functions of interest. This way
of defining probability is applied to various examples and problems including, perhaps most
notably, to a long-running controversy concerning the distinction between Bayesian and fiducial
inference. The characteristics of this definition of probability are carefully evaluated in terms
of the issues that it sets out to address.}
\\[3ex]

\pagebreak
\section{Introduction}

Over the years the issue of how to give the concept of probability a real-world meaning has proved
to be controversial, see for example Fine~(1973), Gillies~(2000) and Eagle~(2011).
Closely related to this issue is the problem of how to adequately elicit subjective probabilities
in any given practical context. Various approaches have been suggested to tackle this latter
problem, see for example, Kadane and Wolfson~(1998), Garthwaite, Kadane and O'Hagan~(2005),
O'Hagan et al.~(2006) and Kynn~(2008).
A procedure incorporated into some of these approaches involves comparing the likeliness of any
given event of interest with the likeliness of various events that each correspond to a given union
of outcomes of a standard experiment, e.g.\ randomly drawing a ball out of an urn containing
distinctly labelled balls, or spinning what is known as a probability wheel (see Spetzler and Stael
von Holstein 1975).

In Bowater~(2017a), a definition of probability was proposed, namely type B probability, that was
based around this type of elicitation procedure, and in particular, on ordering the similarities
that are felt to exist between the likeliness of the event of interest and the likeliness of an
experimental event of the type just mentioned over various specifications for this latter event.
This definition was subsequently extended in Bowater~(2017b) so that continuous probability
distributions could be characterised in an analogous manner to how individual probabilities of
events had been characterised in the original definition, and it was applied to the problem of
statistical inference both in Bowater~(2017b) and Bowater~(2018). 
However, the following criticisms have been raised against this definition of probability:

\vspace*{1ex}
\noindent
1) It is unclear how probabilities can be made to obey the additivity rule of probability, which is
regarded as one of the main aims of the definition.

\vspace*{1ex}
\noindent
2) It is inconvenient that probabilities are only defined at evenly spaced points on the interval
$[0,1]$ with the spacing between points being potentially quite large.

\vspace*{1ex}
\noindent
3) The dependency of probabilities on a reference set of events is unattractive.

\vspace*{1ex}
\noindent
4) The definition does not appear to be universal, e.g.\ the type of characterisation this
definition gives to continuous probability distributions was not extended to probability
distributions for discrete or categorical variables.

\vspace*{1ex}
The main aim of the present paper is to substantially overhaul this definition of probability in a
way that attempts to address these criticisms.
Furthermore, as well as trying to deal with the issue of how the concept of probability can be
given an adequate real-world interpretation, the work outlined in the present paper is motivated,
as to some extent was the case in Bowater~(2017a), by the question of how to explain a particular
type of phenomenon which can not be easily explained by applying the conventional mathematical
definition of probability.
One of the most succinct (but also perhaps one of the least convincing) illustrations of this type
of phenomenon can be found in what is known as Ellsberg's two urn or two colour example, see
Ellsberg~(1961). Even though this is a standard example, some readers may not be familiar with it,
and so it will now be briefly outlined.

Let us imagine that there are two urns that both contain 100 balls, where each ball may be either
red or black in colour. In the first urn, the ratio of red to black balls is entirely unknown,
i.e.\ there may be from 0 to 100 red or black balls in the urn. By contrast, in the second urn it
is known that there are exactly 50 red balls and 50 black balls. An individual is asked to decide
which urn he would prefer to randomly draw a ball out of if getting a red ball wins \$100, while
getting a black ball wins nothing, and which urn he would prefer if a black ball wins \$100, while
a red ball wins nothing.

In Ellsberg~(1961) it is claimed that, first, the majority of people would prefer the second urn in
response to both questions, which is a claim supported by later experiments, e.g.\ Fellner~(1961),
Becker and Brownson~(1964) and Curley and Yates~(1989), and second, and perhaps more importantly,
that having such a preference can not be assumed to be irrational.
However, this type of behaviour is regarded by some as representing a paradox, and is in fact known
as a version of Ellsberg's paradox, as it goes against the idea that an individual would prefer the
urn associated with the highest probability of winning the prize, or be indifferent between urns
that have the same probability of winning the prize.

Having clarified the motivation for this paper, let us give a brief description of its structure.
The main theoretical principles of the definition of probability that will be proposed are laid out
and explained in detail in the sections that immediately follow, in particular, in
Sections~\ref{sec12} to~\ref{sec13}.
A substantive application of this definition of probability is then presented in
Sections~\ref{sec7} and~\ref{sec9}, and a possible extension to the main theory with regard to a
special case is put forward in Section~\ref{sec14}.
The final two sections of the paper discuss how well the theory addresses the questions and issues
that have just been outlined.

\vspace*{3ex}
\section{Fundamental concepts}

\vspace*{1ex}
\subsection{A note about previous definitions}
\label{sec12}

While the theory that will be put forward in the present paper has a great deal in common with the
theory outlined in Bowater~(2017a, 2017b), it is nevertheless a theory that is intended as a
substitute for, rather than an extension of, this earlier theory.
As a result, the definitions used in the present paper generally stand alone from those used in
these earlier papers, and caution is recommended in using this earlier work to try to gain greater
insight about the present work.

\vspace*{2.5ex}
\subsection{Probabilities and probability distributions}
\label{sec2}

In contrast to Bowater~(2017a, 2017b) where the concept of strength was developed separately for
the probability of an event and for continuous probability distributions, here the concept of
strength will be defined primarily as a concept that is applied to distribution functions of
random variables.

A probability distribution will be defined by its (cumulative) distribution function and the
strength of this function relative to other distribution functions of interest.
The distribution function will be defined as having the standard mathematical properties of such a
function. The definition of the concept of strength will be outlined and discussed in detail in
Section~\ref{sec10}, after some more fundamental concepts have been presented in the sections that
immediately follow.

In the theory that will be developed, the probabilities of events will be analysed in the context
of the discrete or continuous distribution functions to which they must be associated. This
includes the simple case where the distribution function is defined by just the probability of a
given event and that of its complement, i.e.\ a Bernoulli distribution function.

\vspace*{2.5ex}
\subsection{Similarity}

Let $S(A,B)$ denote the similarity that a given individual feels there is between his confidence
(or conviction) that an event $A$ will occur and his confidence (or conviction) that an event $B$
will occur. For any three events $A$, $B$ and $C$, it will be assumed that an individual is capable
of deciding whether or not the orderings $S(A,B) > S(A,C)$ and $S(A,B) < S(A,C)$ are applicable.
The notation $S(A,B) = S(A,C)$ will be used to represent the case where neither of these orderings
apply. To clarify, it is not being assumed that $S(A,B)$ and $S(A,C)$ are necessarily numerical
quantities.
Furthermore, for any fourth event $D$, it will not be assumed, in general, that an individual is
capable of deciding whether or not the orderings $S(A,B) > S(C,D)$ and $S(A,B) < S(C,D)$ are
applicable.
Therefore, a similarity $S(A,B)$ can be categorised as a partially orderable attribute of any given
pair of events $A$ and $B$. This is essentially the same definition of the concept of similarity as
used in Bowater~(2017a, 2017b).

\vspace*{3ex}
\subsection{Reference sets of events}
\label{sec1}

\vspace*{1ex}
\noindent
{\bf Definition 1: Discrete reference set of events}

\vspace*{1ex}
\noindent
Let $O=\{ O_1, O_2, \ldots, O_k \}$ be a finite ordered set of $k$ mutually exclusive and
exhaustive events.
It will be assumed that if $O(1)$, $O(2)$ and $O(3)$ are three subsets of the set $O$ that contain
the same number of events, then the following is true:
\vspace*{0.5ex}
\[
S \left( \bigcup_{O_{\hspace*{-0.05em}j} \in\hspace*{0.05em} O(1)} O_{\hspace*{-0.05em}j},
\bigcup_{O_{\hspace*{-0.05em}j} \in\hspace*{0.05em} O(2)} O_{\hspace*{-0.05em}j} \right)
=\, S \left( \bigcup_{O_{\hspace*{-0.05em}j} \in\hspace*{0.05em} O(1)} O_{\hspace*{-0.05em}j},
\bigcup_{O_{\hspace*{-0.05em}j} \in\hspace*{0.05em} O(3)} O_{\hspace*{-0.05em}j} \right)
\vspace*{1.5ex}
\]
for all possible choices of the subsets $O(1)$, $O(2)$ and $O(3)$, where, to clarify, the set
$O(1)$ may be the same as the set $O(2)$ or the set $O(3)$.
Under these assumptions, the discrete reference set of events $R$ is defined by:
\vspace*{-0.5ex}
\begin{equation}
\label{equ1}
R = \{ R(\lambda): \lambda \in \Lambda \}
\vspace*{-0.5ex}
\end{equation}
where $R(\lambda) = O_1 \cup O_2 \cup \cdots \cup O_{\lambda k}$ and
$\Lambda = \{ 1/k,\, 2/k,\, \ldots, (k-1)/k \}$

\vspace*{2ex}
It should be clear that any given individual could easily decide that the set of all the outcomes
of drawing a ball out of an urn containing $k$ distinctly labelled balls could be the set $O$.

\vspace*{2ex}
\noindent
{\bf Definition 2: Continuous reference set of events}

\vspace*{1ex}
\noindent
Let $V$ be a random variable that must take a value in the interval $\Lambda=(0,1)$.
It will be assumed that if $\Lambda(1)$, $\Lambda(2)$ and $\Lambda(3)$ are three subsets of the
interval $(0,1)$ that have the same total length, then the following is true:
\[
S(\{\hspace*{0.05em}V\! \in \Lambda(1)\}, \{\hspace*{0.05em}V\! \in \Lambda(2)\}) =
S(\{\hspace*{0.05em}V\! \in \Lambda(1)\}, \{\hspace*{0.05em}V\! \in \Lambda(3)\})
\]
for all possible choices of the subsets $\Lambda(1)$, $\Lambda(2)$ and $\Lambda(3)$, where, to
clarify, the set $\Lambda(1)$ may be the same as the set $\Lambda(2)$ or the set $\Lambda(3)$.
Under these assumptions, the continuous reference set of events $R$ is defined by
equation~(\ref{equ1}) but now with the set $\Lambda$ defined as it is presently, i.e.\ as the
interval $(0,1)$, and the event $R(\lambda)$ defined to be the event $\{V < \lambda\}$.

\vspace*{2ex}
Let us consider the example of randomly spinning a wheel of unit circumference, and let us assume
that any specific position on the circumference of the wheel is measured as the distance in a given
direction around the circumference from a given point on the circumference.
In such a situation, it should be clear that any given individual could easily decide that the
outcome of a spin of the wheel, as defined by the position on the circumference of the wheel when
it stops that is indicated by a fixed pointer in its centre, could be the variable $V$.

\vspace*{3ex}
\subsection{Scaling events}
\label{sec3}

A scaling event $L(\lambda)$ will be defined as the event $\{V^{*} < \lambda\}$, where $\lambda \in
[0,1]$ and $V^{*}$ has the same definition as the random variable $V$ used in Definition~2 but with
the added condition that it must be the outcome of a well-understood physical experiment, such as
the outcome of spinning the type of wheel described in the previous section.
Since what does or does not constitute a well-understood physical experiment is rather vague, the
definition of a scaling event is open to criticism. The relevance of this criticism should be taken
into account with respect to the way scaling events are used in the rest of this paper.

\vspace*{3ex}
\subsection{Compatibility of reference sets}
\label{sec6}

A discrete or continuous reference set $R_0$ will be defined as being compatible with a discrete or
continuous reference set $R_1$ for all $\lambda \in \Lambda^{\hspace*{-0.05em}*}$ if, first,
$\Lambda^{\hspace*{-0.05em}*} \subset \Lambda_0 \cap \Lambda_1$, where $\Lambda_0$ and $\Lambda_1$
are the sets of allowable values of $\lambda$ for $R_0$ and $R_1$ respectively, and second, it
holds, for all conceivable pairs of events $E_0$ and $E_1$ and all $\lambda \in
\Lambda^{\hspace*{-0.05em}*}$, that
\[
S(E_1,R_{0}(\lambda)) > S(E_0, R_{0}(\lambda))\ \ \ \mbox{if and only if}\ \ \
S(E_1,R_{1}(\lambda)) > S(E_0, R_{1}(\lambda))
\]

For example, we would expect a rational individual to decide that a reference set $R_0$ based on
the outcomes of drawing out a ball from an urn containing 10 distinctly labelled balls is
compatible with a reference set $R_1$ based on the outcomes of drawing out a ball from an urn
containing 100 distinctly labelled balls for all $\lambda \in \Lambda_0 \cap \Lambda_1$, which
means of course for all $\lambda \in \Lambda_0 = \{0.1,0.2,\ldots,0.9\}$.
Also, a rational individual may well decide that a reference set $R_0$ based on the outcomes of
drawing out a ball from an urn containing $k$ distinctly labelled balls is compatible with a
reference set $R_1$ based on the outcome of spinning the type of wheel described in
Section~\ref{sec1} for all $\lambda \in \Lambda_0 \cap \Lambda_1$, i.e.\ for all $\lambda \in
\Lambda_0 = \{1/k,\, 2/k,\, \ldots, (k-1)/k \}$

On the other hand, similar to Ellsberg's two urn example described in the Introduction, let us
imagine that there are two urns that both contain $k$ balls, where each ball has been marked with
a number that is in the range from 1 to $k$. In the first urn, the number of balls that have been
marked with any given number is entirely unknown, i.e.\ there may be 0 to $k$ balls marked with any
given number, while in the second urn, similar to the example that has just been discussed, we know
that there is exactly one ball that has been marked with any given number.
To clarify, we are assuming that there is no particular number in the range 1 to $k$ that we would
regard as being more likely to be on the ball drawn out of the first urn than any other number in
this range.
Here, in comparison to the earlier examples, it would be expected that a much smaller proportion of
rational individuals would be prepared to treat a reference set $R_0$ based on the outcomes of
drawing out a ball from the first urn as being compatible with a reference set $R_1$ based on the
outcomes of drawing out a ball from the second urn for all $\lambda \in \Lambda_0 = \Lambda_1$.

\pagebreak
\section{Strength of a distribution function}
\label{sec10}

\vspace*{1ex}
\subsection{Overview}

As mentioned in Section~\ref{sec2}, in order to complete the definition of a probability
distribution, the strength of its (cumulative) distribution function relative to other distribution
functions of interest needs to be established. In the following sections, the strength of a
distribution function will be defined in terms of the context in which the concept of strength is
being applied.

\vspace*{3ex}
\subsection{When eliciting a distribution function}
\label{sec5}

First, we will define the concept of strength in the case where a given individual is trying to
elicit a distribution function for a given random variable $X$ based on his own personal knowledge.
In this context, it would seem appropriate to use a concept of strength that will be referred to as
internal strength. This concept will now be defined separately for continuous and for discrete
distribution functions.

\vspace*{2ex}
\noindent
{\bf Definition 3: Internal strength for continuous distribution functions}

\vspace*{1ex}
\noindent
Let a given continuous random variable $X$ of possibly various dimensions have two proposed 
distribution functions $F_X(x)$ and $G_X(x)$. We define the set of \vspace*{1ex} events
$\mathcal{F}[a]$ by:
\begin{equation}
\label{equ8}
\mathcal{F}[a] = \left\{ \{ X \in \mathcal{A} \}: \int_{\mathcal{A}} f_X(x) dx = a \right\}\ \ \ \
\mbox{for $a \in [0,1]$}
\vspace*{1.5ex}
\end{equation}
where $\{ X \in \mathcal{A} \}$ is the event that $X$ lies in the set $\mathcal{A}$ and $f_X(x)$ is
the density function corresponding to $F_X(x)$, and let us define the set $\mathcal{G}[a]$ in the
same way with respect to the distribution function $G_X(x)$.
For a given discrete or continuous reference set of events $R$, we will now define the distribution
function $F_X(x)$ \pagebreak as being internally stronger than the distribution function $G_X(x)$
at the resolution level $\lambda$, where $\lambda$ is any value in the set~$\Lambda$ corresponding
to the set $R$, if
\vspace*{-1ex}
\begin{equation}
\label{equ3}
\underset{\mbox{\footnotesize \strut $A\hspace{-0.1em} \in\hspace{-0.05em}
\mathcal{F}[\lambda]$}}{\min}\, S( A, R(\lambda) ) > \underset{\mbox{\footnotesize \strut
$A\hspace{-0.1em} \in\hspace{-0.05em} \mathcal{G}[\lambda]$}}{\min}\, S( A, R(\lambda) )
\vspace*{2.5ex}
\end{equation}

To give an example of the application of this definition, let us imagine that a doctor is trying
to elicit a distribution function for the change $X$ in average survival time that results from the
administration of an untested new drug in comparison to a standard drug. We will assume that the
reference set of events $R$ is based on the outcome of spinning the type of wheel described in
Section~\ref{sec1}, and that the resolution $\lambda$ is some value in the interval $[0.05,0.95]$.
Let $G_X(x)$ be the current proposed distribution function for $X$. The aim is therefore to try to
adjust this distribution function so that it better represents what is known about the variable
$X$, which we will regard as being equivalent to achieving some kind of overall increase in the
similarities in the set $\{ S( A, R(\lambda) ): A \in \mathcal{G}[\lambda]\}$.

In particular, it is natural to put more attention on increasing the lowest similarities in this
set without decreasing by too much, or at all, the highest similarities in this set.
Hence, it would seem sensible to take another step in the elicitation process if an alternative
distribution function $F_X(x)$ is judged as being, according to Definition~3, internally stronger
than the distribution function $G_X(x)$. The distribution function $F_X(x)$ would then become the
current proposed distribution function for $X$, and the elicitation process would continue until no
improvements to this distribution function can be made.

\vspace*{2ex}
\noindent
{\bf Definition 4: Internal strength for discrete distribution functions}

\vspace*{1ex}
\noindent
Let a given discrete random variable $X$ that can only take a value $x$ that belongs to the finite
or countable set $\{ x_1, x_2, \ldots \}$ have two proposed distribution functions $F_X(x)$ and
$G_X(x)$. Also, let the events $L_1(b_1), L_2(b_2), \ldots$ be scaling events \pagebreak (as
defined in Section~\ref{sec3}) that are independent of the variable $X$. Furthermore, we define the
set of events $\mathcal{F}[a]$ by:\linebreak
\vspace*{-0.5ex}
\begin{equation}
\label{equ5}
\mathcal{F}[a] = \left\{\, \bigcup_{i=1}^{\infty}\, (L_i (b_i) \cap \{ X=x_i \})\,
\left|\,\, \sum^{\infty}_{i=1}\, [\, b_i \in (0,1) \,] \leq 1\,\,
\mbox{\Large $\wedge$}\, \sum_{i=1}^{\infty} b_i f_X(x_i)=a \right. \right\}
\end{equation}
\par \vspace*{3ex} \noindent
for $a \in [0,1]$, where $f_X(x)$ is the probability mass function corresponding to $F_X(x)$, and
$[\hspace*{0.2em}]$ on the right-hand side of this equation denotes the indicator function, and let
us define the set of events $\mathcal{G}[a]$ in the same way with respect to $G_X(x)$.
To clarify, all events in the set $\mathcal{F}[a]$ would naturally be assigned a probability of $a$
under the probability mass function $f_X(x)$ for the variable $X$.

For a given discrete or continuous reference set of events $R$, we will now define the function
$F_X(x)$ as being internally stronger than the function $G_X(x)$ at the resolution $\lambda$, where
$\lambda \in \Lambda$, if the condition in equation~(\ref{equ3}) is satisfied with respect to the
definitions currently being used.

\vspace*{2ex}
One of the reasons for the first predicate in the definition of $\mathcal{F}[a]$ in
equation~(\ref{equ5}), i.e.\ the condition that at most only one value in the set $b_1, b_2,
\ldots$ is not equal to 0 or~1, \linebreak is that without this predicate there would be an event
in the set $\mathcal{F}[a]$ that would be effectively equivalent to any of the scaling events
$L_1(a), L_2(a), \ldots$, in particular it would be the event corresponding to setting $b_i=a$
$\forall\hspace*{0.05em} i$.
In other words, there would be an event in this set that would have the undesirable property of
having a definition that does not depend on how the distribution function of interest $F_X(x)$ is
specified.
The practical importance of this issue will perhaps be more clearly seen when this definition of
$\mathcal{F}[a]$ is used again in Section~\ref{sec4}.

The application of Definition~4 of internal strength can be illustrated by imagining that an
election for a state governor has five candidates, and a political analyst is trying to elicit
probabilities for the events $x_1, x_2, \redots, x_5$ of each one of these candidates winning. The
reference set of events $R$ will be specified as in the previous example and again we will assume
that the resolution $\lambda$ is some value in the interval $[0.05,0.95]$.
Also, let the current proposed distribution function and mass function for the variable in question
be $G_X(x)$ and $g_X(x)$ respectively.
At any given stage of the elicitation process, the lowest similarities in the set
$\{ S(A, R(\lambda)) : A \in \mathcal{G}[\lambda] \}$ will usually be caused by one or two of the
probabilities in the set $\{ g_X(x_i) : i=1,2,\redots,5 \}$ being relatively poor representations
of the analyst's beliefs.
This being the case, it would seem natural that the next step in the elicitation process would be
to try to improve the representativeness of these particular probabilities without lowering by too
much, or at all, the representativeness of the other probabilities in the set $\{ g_X(x_i) :
i=1,2,\redots,5 \}$.
Doing this should have the effect of increasing the minimum similarity on the right-hand side of
equation~(\ref{equ3}).
Hence, it would seem sensible to allow an alternative distribution function $F_X(x)$ to replace
$G_X(x)$ as the current proposed distribution function for $X$ if it is, according to Definition~4,
internally stronger than the distribution function $G_X(x)$.

\vspace*{3ex}
\subsection{When comparing elicited or given distribution functions}
\label{sec4}

Although the concept of internal strength can be regarded as the basis of a natural way of
eliciting distribution functions, it does not really provide us with a useful means of comparing
the nature of distribution functions that have been already elicited for different random
variables. Therefore, once a distribution function has been elicited, an alternative concept of
strength is required so that the function can be interpreted in this more outward-looking context.
This alternative concept of strength will be referred to as external strength. It is a concept that
not only can be applied to distribution functions that need to be derived using the kind of
systematic elicitation process referred to in the previous section, but also to distribution
functions that are directly identified as providing the best representations of our beliefs, which
will be referred to as `given' distribution functions, e.g.\ the distribution function of a
variable that represents the outcome of a well-understood physical process. As was the case for
internal strength, the concept of external strength will be defined separately for continuous and
for discrete distribution functions.

\vspace*{2ex}
\noindent
{\bf Definition 5: External strength for elicited or given continuous distribution functions}

\vspace*{1ex}
\noindent
Let two continuous random variables $X$ and $Y$ of possibly different dimensions have elicited or
given distribution functions $F_X(x)$ and $G_Y(y)$ respectively. As in Definition~3, we specify
the set of events $\mathcal{F}[a]$ according to equation~(\ref{equ8}), and let us define the set
$\mathcal{G}[a]$ in the same way but with respect to the variable $Y$ instead of the variable $X$
and the distribution function $G_Y(y)$ instead of $F_X(x)$.

For a given discrete or continuous reference set of events $R$, we will now define the function
$F_X(x)$ as being externally stronger than the function $G_Y(y)$ at the resolution $\lambda$, where
$\lambda \in \Lambda$, if
\vspace*{-1ex}
\begin{equation}
\label{equ4}
\uline{S}_{\hspace*{0.04em}F} = \underset{\mbox{\footnotesize \strut $A\hspace{-0.1em}
\in\hspace{-0.05em} \mathcal{F}[\lambda]$}}{\min}\, S( A, R(\lambda) ) >
\underset{\mbox{\footnotesize \strut $A\hspace{-0.1em} \in\hspace{-0.05em}
\mathcal{G}[\lambda]$}}{\max}\, S( A, R(\lambda) ) =\hspace*{0.05em} \oline{S}_{\hspace*{0.03em}G}
\vspace*{3ex}
\end{equation}

This definition can be interpreted as meaning that if the function $F_X(x)$ is judged as being
externally stronger than the function $G_Y(y)$ then, relative to the reference event $R(\lambda)$,
it better represents the uncertainty associated with the variable $X$ than $G_Y(y)$ represents the
uncertainty associated with the variable $Y$.

In comparison to the definition of internal strength in equation~(\ref{equ3}), it is naturally
appealing to have the maximization operator on the right-hand side of equation~(\ref{equ4}) instead
of the minimization operator, as this of course implies that all the similarities in the set
$\{ S( A, R(\lambda) ) : A \in \mathcal{F}[\lambda] \}$ are greater than any similarity in the set
$\{ S( A, R(\lambda) ) : A \in \mathcal{G}[\lambda] \}$. However, it would not have been sensible
to have defined internal strength such that the maximization instead of the minimization operator
appears on the right-hand side of equation~(\ref{equ3}), as using such a strong condition as the
basis for a probability elicitation process would generally impede the ease with which such a
process could develop.

To give an example of the application of Definition~5, let us compare a uniform distribution
function $F_X(x)$ over $(0,1)$ for the output $X$ of a pseudo-random number generator that has been
carefully designed to produce approximately uniform random numbers in $(0,1)$ with a doctor's
elicited distribution function $G_Y(y)$ for the change $Y$ in average survival time caused by the
administration of an untested new drug in comparison to a standard drug. The reference set of
events $R$ and the range of the resolution $\lambda$ will be specified as in the examples of
Section~\ref{sec5}.

Under these assumptions, all the similarities in the set
$\{ S( A, R(\lambda) ) : A \in \mathcal{F}[\lambda] \}$ may well be regarded as being quite high.
This is because the event $R(\lambda)$ is the outcome of a well-understood physical experiment,
i.e.\ a random spin of a wheel, while any event in the set $\mathcal{F}[\lambda]$ may well feel
like it can be almost treated as though it is the outcome of a well-understood physical experiment.
On the other hand, the doctor's uncertainty about whether or not any given event in the set
$\mathcal{G}[\lambda]$ will occur could be regarded as depending largely on his incomplete
knowledge about highly complex biological processes in the human body.
Therefore, it could reasonably be expected that, according to Definition~5, the doctor would
consider the function $F_X(x)$ as being externally stronger than the function $G_Y(y)$, which if
being the case, could be interpreted as meaning that, relative to the spin-of-a-wheel event
$R(\lambda)$, he feels that the function $F_X(x)$ performs better than the function $G_Y(y)$ at
representing the uncertainty that these functions are intended to represent.

\vspace*{2ex}
\noindent
{\bf Definition 6: External strength for elicited or given discrete distribution\\ functions}

\vspace*{1ex}
\noindent
Let $X$ and $Y$ be two discrete random variables that can only take values in the finite or
countable sets $x=\{ x_1, x_2, \ldots \}$ and $y=\{ y_1, y_2, \ldots \}$ respectively, and let
$F_X(x)$ and $G_Y(y)$ be elicited or given distribution functions for these two variables
respectively. Also, let the events $L_1(b_1), L_2(b_2), \ldots$ be scaling events that are
independent of the variables $X$ and $Y$.
Furthermore, as in Definition~4, we assume the set of events $\mathcal{F}[a]$ is specified
according to equation~(\ref{equ5}), and let us define the set $\mathcal{G}[a]$ in the same way but
with respect to the variable $Y$ and the distribution function $G_Y(y)$.

For a given discrete or continuous reference set of events $R$, we will now define the function
$F_X(x)$ as being externally stronger than the function $G_Y(y)$ at the resolution $\lambda$, where
$\lambda \in \Lambda$, if the condition in equation~(\ref{equ4}) is satisfied with respect to the
definitions currently being used.

\vspace*{2ex}
This definition can be applied to the motivating example referred to in the Introduction, i.e.\
Ellsberg's two urn example. In particular, we will denote the outcomes of drawing a ball out of the
first urn and the second urn in this example as the random variables $X$ and $Y$ respectively, and
we will denote the distribution functions for these two variables as $F_X(x)$ and $G_Y(y)$
respectively.
The reference set $R$ and the range of $\lambda$ will again be specified as in the examples of
Section~\ref{sec5}.
Furthermore, let us imagine that, with regard to both the first and the second urns, an individual
elicits a probability mass function that assigns a probability of 0.5 to both of the potential
outcomes of drawing out a ball from the urn concerned, i.e.\ the outcomes of drawing out a red ball
and drawing out a black ball.
This would seem to be quite a rational decision to make.

Since both $F_X(x)$ and $G_Y(y)$ effectively define a Bernoulli distribution function, the sets of
events $\mathcal{F}[\lambda]$ and $\mathcal{G}[\lambda]$ will each only contain two events.
For example, the set $\mathcal{F}[0.5]$ simply contains the events of drawing out a red ball and
drawing out a black ball from the first urn, while the set $\mathcal{G}[0.5]$ contains the same two
events but with respect to the second urn.
Now, given the ambiguity surrounding the uncertainty about whether or not any given one of the two
events in $\mathcal{F}[0.5]$ will occur, it should be fairly clear why the individual is likely to
decide that the similarities between the spin-of-a-wheel event $R(0.5)$ and the events in
$\mathcal{G}[0.5]$ are higher than the similarities between $R(0.5)$ and the events in
$\mathcal{F}[0.5]$.
According to Definition~6, by doing this, he would be of course deciding that the distribution
function $G_Y(y)$ is externally stronger than the distribution function $F_X(x)$ at a resolution
level of 0.5.
A similar line of reasoning can be used to justify the individual drawing the same conclusion with
respect to other values for the resolution $\lambda$.
Therefore if, all other things equal, the individual would prefer being in the scenario in
Ellsberg's two urn example where the outcomes of winning \$100 and of winning nothing are outcomes
over which he has elicited a distribution function that is externally stronger than the
distribution function that he has elicited over the same outcomes in the alternative scenario, then
the `paradoxical' behaviour identified by Ellsberg in this example can be accounted for by the
definition of probability outlined in the present work.

We could also apply Definition~6 of external strength to the governor election example outlined in
Section~\ref{sec5} under the assumption that the political analyst has already used Definition~4 of
internal strength to elicit a distribution function $H_Z(z)$ over the events of each of the five
candidates winning, which will now be denoted as the events $z_1, z_2, \redots, z_5$.
Given that the factors that can influence which candidate is elected are likely to be considered
vague and difficult to weigh up, it should be fairly clear, under the same assumptions about the
set $R$ and the range of $\lambda$ as we have been making, why the analyst is likely to consider
his distribution function $H_Z(z)$ as being externally weaker than the distribution function
$G_Y(y)$ from Ellsberg's two urn example.
However, it would be much less easy to predict whether any given political analyst in the situation
of interest would decide that his distribution function $H_Z(z)$ is externally stronger, weaker or
neither stronger nor weaker than the distribution function $F_X(x)$ from this earlier example.

\pagebreak
\subsection{Sensitivity to the choice of the reference set \texorpdfstring{$R$}{R} and the
resolution \texorpdfstring{$\lambda$}{lambda}}
\label{sec8}

In general, Definitions~3 to~6 of internal and external strength depend on the reference set of
events $R$ being used. More comments will be made with regard to this matter in
Section~\ref{sec11}. However for now, let us clarify that if, according to the definition given in
Section~\ref{sec6}, a discrete or continuous reference set $R_0$ is compatible with another
discrete or continuous reference set $R_1$ for all $\lambda \in \Lambda^{\hspace*{-0.05em}*}$, then
Definitions~3 to~6 will not be affected by whether the reference set $R_0$ or $R_1$ is used,
provided that the resolution $\lambda \in \Lambda^{\hspace*{-0.05em}*}$.

With regard to the choice of the resolution level $\lambda$, it is known that people have
difficulty in weighing up the uncertainty associated with the occurrence of events that are very
unlikely or very likely to occur, which is a disadvantage that could apply if $\lambda$ was less
than say 0.05 or greater than say 0.95.
On the other hand, it could be argued that the further that $\lambda$ is away from the value 0.5,
the greater the detail in which the characteristics of the distribution functions involved in the
Definitions~3 to~6 may be explored.
Nevertheless, it would be expected that, in many applications, Definitions~3 to~6 will be largely
insensitive to the choice made for the value of $\lambda$ over the range $[0.05, 0.95]$, which is
the range for $\lambda$ that has been used in the examples that have been considered so far.

\vspace*{3ex}
\subsection{When comparing distribution functions derived by formal reasoning}
\label{sec13}

Of course, not all distribution functions can be regarded as having been derived by some method of
direct evaluation. 
Therefore, let us now turn our attention to defining the concept of strength in the case where we
wish to compare the nature of distribution functions that have been derived using any type of
procedure, including through the use of a formal system of reasoning, e.g.\ derived by applying the
standard rules of probability.
In particular, this will be achieved by simply using a more general definition of the concept of
external strength than the definitions of this concept presented in Section~\ref{sec4}.

\pagebreak
\noindent
{\bf Definition 7: General definition of external strength}

\vspace*{1ex}
\noindent
Let two random variables $X$ and $Y$ have distribution functions $F_X(x)$ and $G_Y(y)$
respectively. To clarify, no assumption is being made about whether the variables $X$ and $Y$ are
discrete or continuous, and indeed one of these variables may be continuous, while the other one
may be discrete.
Also, let $\mathcal{M}_F$ and $\mathcal{M}_G$ be two sets of reasoning processes that could be used
to evaluate the minimum similarity $\uline{S}_{\hspace*{0.04em}F}$ and the maximum similarity
$\oline{S}_{\hspace*{0.03em}G}$ respectively, where these similarities are as defined in
equation~(\ref{equ4}), and where the sets $\mathcal{F}[\lambda]$ and $\mathcal{G}[\lambda]$
referred to in this equation are defined according to Definition~5 or Definition~6 depending on
whether the random variable on which the set is based is continuous or discrete.

For a given discrete or continuous reference set of events $R$, the function $F_X(x)$ will now be
defined as being externally stronger than the function $G_Y(y)$ at the resolution $\lambda$, where
$\lambda \in \Lambda$, if
\begin{equation}
\label{equ7}
\underset{\mbox{\footnotesize \strut $M\hspace{-0.1em} \in \hspace{-0.05em}
\mathcal{M}_F$}}{\max}\,\, \uline{S}_{\hspace*{0.04em}F}\,\, >\,
\underset{\mbox{\footnotesize \strut $M\hspace{-0.1em} \in \hspace{-0.05em}
\mathcal{M}_G$}}{\max}\hspace*{0.25em} \oline{S}_{\hspace*{0.03em}G}
\vspace*{2ex}
\end{equation}
where $M \in \mathcal{M}$ denotes `over all reasoning processes in the set $\mathcal{M}$'.

\vspace*{2ex}
It is natural to regard Definitions~5 and~6 of external strength as corresponding to special cases
of this definition in which the variables $X$ and $Y$ are either both continuous or both discrete,
and in which the sets $\mathcal{M}_F$ and $\mathcal{M}_G$ each contain only one reasoning process,
which is the method of direct evaluation.
More generally, though, we are faced with the problem that the definition of external strength just
presented may depend on the choices that are made for the sets $\mathcal{M}_F$ and $\mathcal{M}_G$.
In many cases, this problem can be avoided to a great extent by choosing the sets $\mathcal{M}_F$
and $\mathcal{M}_G$ to be large enough so that they arguably contain all methods of reasoning that
are relevant to evaluating the similarities concerned.
However, as will be illustrated in Section~\ref{sec7}, this may be difficult to achieve if there
are one or more potentially relevant methods of reasoning that are not well understood.

Observe that when distribution functions are derived by formal systems of reasoning rather than by
a direct method of evaluation, the problem also arises that the distribution function for any given
random variable may itself depend on which system of reasoning is used to derive it. Due to this
possibility, the following definition is required.

\vspace*{2ex}
\noindent
{\bf Definition 8: Criterion for choosing between distribution functions}

\vspace*{1ex}
\noindent
Let $F_X(x)$ and $G_X(x)$ be two proposed distribution functions for the random variable $X$ that
have been derived using two separate methods of reasoning.
Now, if the random variable $Y$ in Definition~7 is assumed to be equivalent to $X$, and the sets
$\mathcal{M}_F$ and $\mathcal{M}_G$ in this definition are regarded by the given individual who has
the task of evaluating the similarities in equation~(\ref{equ7}) as containing all methods of
reasoning that are relevant for this task, then the function $F_X(x)$ will be favoured over
$G_X(x)$ as being the distribution function of $X$ if it is externally stronger than $G_X(x)$
according to Definition~7.

\vspace*{2ex}
We can interpret this definition as meaning that $F_X(x)$ will be favoured over $G_X(x)$ as being
the distribution function of $X$ if, relative to the reference event $R(\lambda)$, it better
represents the uncertainty associated with the variable $X$ than the function $G_X(x)$.

\vspace*{3ex}
\subsection{Applying the concept of strength to the
Bayesian\hspace*{0.03em}-\hspace*{0.03em}fiducial controversy}
\label{sec7}

In this section, we will apply the concept of external strength to the controversy about whether
fiducial reasoning is of any use in circumstances where the fiducial density function is equal to a
posterior density function corresponding to a given choice of the prior density function.
We will deliberately restrict our attention to the case where inferences need to be made about the
mean $\mu$ of a normal density function that has a known variance $\sigma^2$ on the basis of a
random sample $x$ of $n$ values drawn from the density function in question, since it will be seen
that the issues that are explored in analysing this case are relevant to many other cases.
The type of fiducial inference that will be applied will be subjective fiducial inference as
outlined in Bowater~(2018).

Let it be assumed that very little or nothing was known about $\mu$ before the sample $x$ was
observed. In a Bayesian analysis, it would be quite conventional to try to represent this lack of
knowledge by placing a diffuse symmetric prior density over $\mu$ centred at some given value for
its median $\mu_0$.
Assuming this has been done, let the corresponding prior and posterior distribution functions be
denoted as $D(\mu)$ and $D(\mu\,|\,x)$ respectively.
However, these distribution functions are not sufficient to define the prior and posterior
distributions of $\mu$ under the definition of probability being considered in the present paper.
As was explained in Section~\ref{sec2}, to complete the definitions of these probability
distributions, we need to evaluate the strengths of their corresponding distribution functions
relative to other distribution functions of interest.
In the current context, this clearly needs to be done by applying the concept of external rather
than internal strength, and since it is also clear that, with respect to the distribution function
$D(\mu\,|\,x)$, this needs to be done by applying Definition~7 of external strength, we will use
this general definition of external strength to carry out this task with respect to both the
distribution functions in question.

In particular, in applying this definition, it will be assumed that there is only the method of
direct evaluation in the set of reasoning processes $\mathcal{M}_{\mbox{\footnotesize $\mu$}}$ that
will used to evaluate the similarities in equation~(\ref{equ7}) with respect to the prior
distribution function $D(\mu)$, and that there is only Bayesian reasoning in the set of reasoning
processes $\mathcal{M}_{\mbox{\footnotesize $\mu|x$}}$ that will used to evaluate these
similarities with respect to the posterior distribution function $D(\mu\,|\,x)$.
By Bayesian reasoning, it is meant any system of reasoning that is related to the way that Bayes'
theorem updates the prior to the posterior distribution function by combining it with the
likelihood function. The reference set $R$ and the range of $\lambda$ will be again specified as in
the examples of Section~\ref{sec5}.

Under these assumptions, if the set of events $\mathcal{D}_{\mbox{\footnotesize $\mu$}}[a]$ is
defined as the set $\mathcal{F}[a]$ was de\-fined in equation~(\ref{equ8}) but with respect to the
variable $\mu$ rather than the variable $X$, and the prior distribution function $D(\mu)$ rather
than the generic distribution function $F_X(x)$, then it would be expected that the similarities in
the set $\{ S( A, R(\lambda) ) : A \in \mathcal{D}_{\mbox{\footnotesize $\mu$}}[\lambda] \}$ would
all be regarded as being very low.
In fact, we would expect that it would be difficult, if not impossible, to find any directly
elicited distribution function, for any random variable in any context, that could be regarded as
being externally weaker than the prior distribution function $D(\mu)$ according to Definition~7.
This is because, apart from needing to satisfy the condition that it is diffuse and symmetric, the
choice of the prior density function for $\mu$ when there is very little or no prior information
about $\mu$ will be extremely arbitrary, implying that the definition of the events in the set
$\mathcal{D}_{\mbox{\footnotesize $\mu$}}[\lambda]$ will be just as arbitrary.
For example, if $\lambda=0.5$ then the set $\mathcal{D}_{\mbox{\footnotesize $\mu$}}[\lambda]$ will
contain the events $\{\mu < \mu_0\}$ and $\{\mu > \mu_0\}$ which clearly depend on the very
arbitrary choice of the prior median $\mu_0$.
A similar point about how the choice of the location of the prior density of $\mu$ can affect the
definition of the events in the set $\mathcal{D}_{\mbox{\footnotesize $\mu$}}[\lambda]$ can be made
with respect to other values of $\lambda$ in the range of interest, i.e.\ $[0.05, 0.95]$.
Furthermore, for all values of $\lambda$ in this range, the events in the set
$\mathcal{D}_{\mbox{\footnotesize $\mu$}}[\lambda]$ will of course also generally depend on the
arbitrary decision that needs to be made about how diffuse the prior density for $\mu$ should be
over the real line.

Now, similar to how the set $\mathcal{D}_{\mbox{\footnotesize $\mu$}}[a]$ was defined, let the set
$\mathcal{D}_{\mbox{\footnotesize $\mu|x$}}[a]$ be defined as the set $\mathcal{F}[a]$ was defined
in equation~(\ref{equ8}) but with respect to the variable $\mu$ and the posterior distribution
function $D(\mu\,|\,x)$.
Since the posterior density function for $\mu$ is determined through Bayes' theorem simply by
reweighting the prior density function for $\mu$, that is, by normalising the density function that
results from multiplying this prior density function by the likelihood function of $\mu$, it would
seem difficult to apply a Bayesian reasoning process, i.e.\ a member of the set of reasoning
processes $\mathcal{M}_{\mbox{\footnotesize $\mu|x$}}$, to argue that the similarities in the set
$\,\{\, S( A, R(\lambda) ) : A \in \mathcal{D}_{\mbox{\footnotesize
$\mu|x$}}[\lambda] \,\}\,$ should be generally that much larger than the similarities in the set
$\{\, S( A, R(\lambda) ) : A \in \mathcal{D}_{\mbox{\footnotesize $\mu$}}[\lambda] \,\}$, assuming
that these latter similarities are evaluated before the data are observed.
For this reason, under the assumptions that have been made, it could be argued that, except
possibly for distribution functions elicited in a similar context to how $D(\mu)$ was elicited, it
should not be easy to find a directly elicited distribution function in any circumstances that
could be regarded as being externally weaker than the posterior distribution function
$D(\mu\,|\,x)$ according to Definition~7.

Observe that, in the situation of interest, it would be considered part of common practice to try
to approximate the posterior distribution function $D(\mu\,|\,x)$ with a distribution function
$C(\mu\,|\,x)$ that is the result of using Bayes' theorem to update a prior density function of the
form $c(\mu)=\mbox{constant}$ $\forall \mu \in (-\infty, \infty)$.
Let us first point out that, under the definition of probability being used in the present work, it
would seem inappropriate to refer to $C(\mu\,|\,x)$ as a posterior distribution function, since it
is based on a prior density function $c(\mu)$ that does not follow the standard mathematical rules
of probability, in particular it is an improper density function. Second, since the function
$C(\mu\,|\,x)$ is only being used to approximate the function $D(\mu\,|\,x)$, the external strength
of this function relative to other distribution functions is naturally inherited from
$D(\mu\,|\,x)$, i.e.\ it must be roughly equivalent, in this context, to the external strength that
is assigned to $D(\mu\,|\,x)$ relative to other distribution functions.

Let us now turn our attention to the application of subjective fiducial inference to the case of
interest. The terminology that will be used corresponds to Bowater~(2018), nevertheless the way
that subjective fiducial inference will be applied to this case is equivalent to what was outlined
in both Bowater~(2017b) and Bowater~(2018).

Since the sample mean $\bar{x}$ is a sufficient statistic for $\mu$, it can be assumed to be the
\emph{fiducial statistic} in this case. For the sake of argument, let us also suppose that the
\emph{primary random variable} (primary r.v.) $\Gamma$ has a standard normal density. Making these
assumptions effectively implies that it is being assumed that the data set $x$ was generated by the
following \emph{data generating algorithm}:

\vspace*{1ex}
\noindent
1) Generate a value $\gamma$ for the primary r.v.\ $\Gamma$ by randomly drawing this value from the
standard normal density function.

\vspace*{1ex}
\noindent
2) Determine the observed sample mean $\bar{x}$ by setting $\Gamma$ equal to $\gamma$ and
$\xoline{X}$ equal to $\bar{x}$ in the following expression:
\vspace*{0.5ex}
\begin{equation}
\label{equ9}
\xoline{X} = \mu+(\sigma/\sqrt{n}\hspace*{0.1em})\hspace*{0.05em}\Gamma
\vspace*{0.5ex}
\end{equation}
which effectively defines the distribution function of the unobserved sample mean $\xoline{X}$.

\vspace*{1ex}
\noindent
3) Generate the data set $x$ by conditioning the joint density function of this data set given
$\mu$ and $\sigma$ on the already generated value of the sample mean $\bar{x}$.

\vspace*{1ex}
It now follows that the fiducial distribution function of $\mu$ is defined, according to the
general rule given in Bowater~(2018), by setting $\xoline{X}$ equal to $\bar{x}$ in
equation~(\ref{equ9}) and then treating the value $\mu$ in this equation as being a random
variable, which implies that this distribution function is alternatively specified by the
expression:
\begin{equation}
\label{equ11}
\mu\, |\, \sigma^2, x \sim \mbox{N}\hspace*{0.05em} (\bar{x}, \sigma^2/n)
\vspace*{-0.5ex}
\end{equation}
The validity of this distribution function as a post-data distribution function for $\mu$ clearly
depends on the argument that the distribution function of the primary r.v.\ $\Gamma$ after the data
$x$ were observed, i.e.\ the post-data distribution function of $\Gamma$, should be the same as the
distribution function of $\Gamma$ before the data $x$ were observed.
As far as this discussion is concerned, this argument will be regarded as being the fiducial
argument.

The fiducial distribution function of $\mu$ defined by equation~(\ref{equ11}) is the same as the
distribution function $C(\mu\,|\,x)$ that was defined earlier.
However, to assess the external strength of this distribution function relative to other
distribution functions, it will now be assumed that fiducial reasoning is the only type of
reasoning \pagebreak in the set of reasoning processes $\mathcal{M}_C$ that, in keeping with
Definition~7, will be used to evaluate the similarities in the set
$\{ S( A, R(\lambda)): A \in \mathcal{C}_{\mbox{\footnotesize $\mu|x$}}[\lambda] \}$, where the set
$\mathcal{C}_{\mbox{\footnotesize $\mu|x$}}[a]$ is defined as the set $\mathcal{F}[a]$ was defined
in equation~(\ref{equ8}) but with respect to the variable $\mu$ and the distribution function
$C(\mu\,|\,x)$.
By fiducial reasoning it is meant any system of reasoning that directly attempts to justify the
fiducial argument that was just described.

To help us make the assessment of external strength in question, let us re-analyse one of the
abstract examples that were outlined in Bowater~(2017b).
In the example of interest, it is imagined that someone, who will be referred to as the selector,
randomly draws a ball out of an urn containing seven red balls and three blue balls and then,
without looking at the ball, hands it to an assistant. The assistant, by contrast, looks at the
ball, but conceals it from the selector, and then places it under a cup. The selector believes that
the assistant smiled when he looked at the ball. Finally, the selector is asked to assign a
probability to the event that the ball under the cup is red.
We will assume that it is uncertain whether the assistant knew from the outset that the selector
would be asked to assign a probability to this particular event.

In this scenario, let us now imagine that, relative to other distribution functions of interest,
the selector wishes to evaluate the external strength of the Bernoulli distribution function
$B_Y(y)$ that corresponds to assigning a probability of 0.7 to the event that the ball under the
cup is red ($y=1$), and a probability of 0.3 to the event that it is blue ($y=0$).
This means that, if the set $\mathcal{B}[a]$ is defined as the set $\mathcal{F}[a]$ was defined in
equation~(\ref{equ5}) but with respect to the variable $Y$ and the distribution function $B_Y(y)$,
then the selector will need to evaluate the similarities in the set $\{ S( A, R(\lambda)):
A \in \mathcal{B}[\lambda] \}$, which of course will contain only two similarities as
$\mathcal{B}[\lambda]$ can only contain two events.

In doing this, it will be assumed that the selector takes into account the fact that a smile by the
assistant would be information that could imply that it is less likely or more likely that the ball
under the cup is red. Therefore, his evaluation of the similarities in question must depend on his
subjective judgement regarding the meaning of the assistant's supposed smile. Nevertheless, he may
feel that, if the assistant had indeed smiled, he would not really have understood the smile's
meaning. In this case, it would seem rational for him to conclude that the similarities in the set
$\{ S( A, R(\lambda)): A \in \mathcal{B}[\lambda] \}$ could be at least approximately evaluated by
making the assumption that, without looking at the ball, he had placed it directly under the cup,
rather than giving the assistant an opportunity to look at the ball.
Under this assumption, since along with the event $R(\lambda)$, the propensity of either of the
two events in $\mathcal{B}[\lambda]$ to occur would only depend on the outcome of a well-understood
physical experiment, it would be expected that he would regard both of the similarities in the set
$\{ S( A, R(\lambda)): A \in \mathcal{B}[\lambda] \}$ as being equal or very close to the highest
possible similarity that can exist between two events.
Therefore, the same conclusion could be justified as being valid or approximately valid in the
original scenario where the assumption in question is not made.

Returning to the evaluation of the relative external strength of the fiducial distribution function
$C(\mu\,|\,x)$, let us assume that, in step~1 of the data generating algorithm outlined above, the
value $\gamma$ of the primary r.v.\ $\Gamma$ is generated by a well-understood physical experiment,
which is usually a very reasonable assumption to make.
We will now make an analogy between the uncertainty about the value of $\gamma$ after the data have
been observed in this case, and the uncertainty about the colour of the ball under the cup in the
scenario that has just been outlined.
In particular, given that very little or nothing was known about $\mu$ before the data $x$ were
observed, the event of observing the data should be akin to the event of the selector believing
that the assistant smiled when he looked at the ball in question, and hence this event should have
little or no effect on the nature of the uncertainty that is felt about the value of $\gamma$.

As a result if, the post-data distribution function of $\Gamma$ is chosen to be equal to the
distribution function of $\Gamma$ before the data were observed, i.e.\ equal to a standard normal
distribution function, then it would be expected that, when evaluated after the data have been
observed, the relative external strength of this distribution function of $\Gamma$ would be
regarded as being similar to the relative external strength of the function $B_Y(y)$.
Since, as has already been illustrated, the function $C(\mu\,|\,x)$ can be regarded as being a
post-data distribution function of $\mu$ that, using just simple mathematics, is fully defined by
the post-data distribution function of $\Gamma$ in question and known constants, it can therefore
be argued that the similarities in the set
$\{ S( A, R(\lambda)): A \in \mathcal{C}_{\mbox{\footnotesize $\mu|x$}}[\lambda] \}$ as defined
earlier should all be regarded as being equal or quite close to the highest possible similarity
that can exist between two events.

This conclusion could hardly be more different to the conclusion that was reached earlier when the
relative external strength of the same distribution function $C(\mu\,|\,x)$ was evaluated by
effectively taking into account its approximation to the distribution function $D(\mu\,|\,x)$, and
then applying only Bayesian reasoning.
Therefore, in accordance with the definition of a probability distribution given in 
Section~\ref{sec2}, it could be argued that the posterior distribution for $\mu$ that results from
the use of a flat improper prior density for $\mu$ would be better described as the fiducial
distribution for $\mu$, since, after the data have been observed, the relative external strength of
its corresponding distribution function, i.e.\ $C(\mu\,|\,x)$, is naturally justified as being much
higher by using fiducial rather than Bayesian reasoning.
Moreover, it could be argued that it is possible to justify the relative external strength of the
function $C(\mu\,|\,x)$ as being higher by using fiducial reasoning than by using any other
currently known form of statistical reasoning.

\vspace*{3ex}
\subsection{Example of using Definition~8 to choose between distribution functions}
\label{sec9}

Let us now illustrate how the criterion in Definition~8 can be applied to determine which out of
two possible distribution functions for a given random variable is the most appropriate
distribution function for that variable.
In particular, we will imagine that in the case analysed in the previous section, there is now a
notable, but nevertheless, a fairly low level of prior belief that $\mu$ will not be a very long
distance from a given value $\mu_1$. It will be assumed that in applying the Bayesian method, this
prior belief about $\mu$ is represented by a normal prior density function for $\mu$ with mean
$\mu_1$ and a moderate to large variance.
Let the resulting posterior distribution function be denoted by $I(\mu\,|\,x)$, and also let the
set $\mathcal{I}_{\mbox{\footnotesize $\mu|x$}}[a]$ be defined as the set $\mathcal{F}[a]$ was
defined in equation~(\ref{equ8}) but with respect to the variable $\mu$ and the distribution
function $I(\mu\,|\,x)$.
Alternatively, we could apply the fiducial method to this problem under the assumption that it may
be adequate to not actually take into account the prior belief about $\mu$ just mentioned in
forming a post-data distribution function for $\mu$. Therefore, again the fiducial distribution
function for $\mu$ will be $C(\mu\,|\,x)$ as defined in the last section.

We will now apply Definition~8 to choose which is the most appropriate distribution function for
$\mu$ out of $I(\mu\,|\,x)$ and $C(\mu\,|\,x)$ after the data have been observed. In doing this,
let us assume that the set of reasoning processes that will be used to evaluate both the set of
similarities $\{ S( A, R(\lambda)): A \in \mathcal{I}_{\mbox{\footnotesize $\mu|x$}}[\lambda] \}$
and the set $\{ S( A, R(\lambda)): A \in \mathcal{C}_{\mbox{\footnotesize $\mu|x$}}[\lambda] \}$
contains both Bayesian and fiducial reasoning (as defined in the previous section) but no other
method of reasoning. It is clear though that the former set of similarities can only be evaluated
indirectly using fiducial reasoning, while the latter set of similarities can only be evaluated
indirectly using Bayesian reasoning.

If we apply Bayesian reasoning to evaluate the similarities $\{ S( A, R(\lambda)): A \in
\mathcal{I}_{\mbox{\footnotesize $\mu|x$}}[\lambda] \}$, then since choosing a prior density
function to represent the prior beliefs about $\mu$ in question is still fairly arbitrary, it would
be expected that these similarities will be regarded in general as being only moderately higher
than the similarities in the set $\{ S( A, R(\lambda) ) : A \in \mathcal{D}_{\mbox{\footnotesize
$\mu|x$}}[\lambda] \}$, under the assumption that these latter similarities are evaluated using
Bayesian reasoning in the context of the case considered in the last section.

To evaluate the similarities $\hspace*{0.2em}\{\hspace*{0.2em} S( A, R(\lambda)): A \in
\mathcal{C}_{\mbox{\footnotesize $\mu|x$}}[\lambda] \hspace*{0.2em}\}\hspace*{0.2em}$ in the
present context by applying fiducial reasoning, it again seems sensible to first analyse the
relative external strength of the standard normal distribution function as the distribution
function of the primary r.v.\ $\Gamma$ after the data have been observed.
In particular, it would seem without doubt that the presence of the prior beliefs about $\mu$ in
question should have the effect of lessening our degree of comfort, compared to the case examined
in the previous section, in assuming that this post-data distribution function of $\Gamma$ is a
standard normal distribution function.
However, since the prior beliefs about $\mu$ have been assumed to be fairly vague, this effect may
be considered to be quite small even if the sample size $n$ is small.
Therefore, although the similarities in the set $\{\hspace*{0.1em} S( A, R(\lambda)): A \in
\mathcal{C}_{\mbox{\footnotesize $\mu|x$}}[\lambda] \hspace*{0.1em}\}$ may be regarded in general
as being lower than if these similarities were evaluated using fiducial reasoning in the context of
the case discussed in the last section, they nevertheless may be regarded in general as being
higher than the similarities in the set $\{ S( A, R(\lambda)): A \in
\mathcal{I}_{\mbox{\footnotesize $\mu|x$}}[\lambda] \}$, assuming that these latter similarities
are evaluated either directly by using Bayesian reasoning or indirectly by using fiducial
reasoning.
To clarify, the indirect use of fiducial reasoning to evaluate the similarities
$\{ S( A, R(\lambda)): A \in \mathcal{I}_{\mbox{\footnotesize $\mu|x$}}[\lambda] \}$ would
naturally involve treating the distribution function $I(\mu\,|\,x)$ as being an approximation to
$C(\mu\,|\,x)$, which of course would be expected to have a negative impact on the evaluation of
the similarities concerned in comparison to this approximation not being used.

The conclusion that has just been made could be interpreted as meaning that according to
Definition~7, the distribution function $C(\mu\,|\,x)$ would be regarded as being externally
stronger than the function $I(\mu\,|\,x)$ for values of the resolution $\lambda$ in the range of
interest, i.e.\ [0.05,0.95], which in turn would imply that, according to Definition~8, the
function $C(\mu\,|\,x)$ would be favoured over $I(\mu\,|\,x)$ as being the distribution function of
$\mu$ after the data have been observed.
The case has therefore been made in this section that, if Bayesian and fiducial reasoning are the
only allowable reasoning processes in evaluating the similarities of interest, then even when there
is notable prior information about $\mu$, it may not in fact be worth taking into account this
information in establishing a post-data distribution function for $\mu$, due to the detrimental
effect this may have on the relative external strength of the distribution function concerned.
This may be considered to be somewhat of a paradox.

\vspace*{3ex}
\subsection{Continuous measurement of external strength}
\label{sec14}

The concept of external strength has been defined as an ordinal measurement, i.e.\ using the
definitions that have been given we are able to rank distribution functions in terms of their
external strength. However, a question that naturally arises is whether it is possible to measure
on a continuous scale some kind of characteristic that incorporates the essence of the concept of
external strength.
In this section, we will consider how this could be done in the special case where we wish to
compare the external strengths of distribution functions that have already been classified as
even-similarity distribution functions according to the following definition.

\vspace*{2ex}
\noindent
{\bf Definition 9: An even-similarity distribution function}

\vspace*{1ex}
\noindent
For a given discrete or continuous reference set of events $R$, we will define the distribution
function $F_X(x)$ of the random variable $X$ as being an even-similarity distribution function at
the resolution level $\lambda$, if the minimum similarity $\uline{S}_{\hspace*{0.04em}F}$ is equal
to the maximum similarity $\oline{S}_{\hspace*{0.03em}F}$, where the definitions of these
similarities are as given in equation~(\ref{equ4}), and where the assumptions underlying this
equation correspond to Definition~5 if the variable $X$ is continuous or to Definition~6 if this
variable is discrete.

\vspace*{2ex}
\noindent
A continuous measure of external strength could now be defined in the following way.

\vspace*{2ex}
\noindent
{\bf Definition 10: Proposed continuous measure of external strength}

\vspace*{1ex}
\noindent
With respect to a given reference set of events $R$ and a given resolution $\lambda$, let $F_X(x)$,
$G_Y(y)$ and $H_Z(z)$ be even-similarity distribution functions of any three given random variables
$X$, $Y$ and $Z$ respectively, where according to Definition~7, the function $G_Y(y)$ is not
externally weaker than $F_X(x)$ and is not externally stronger than $H_Z(z)$.
We define the set $\mathcal{F}[\lambda]$ according to Definition~3 or Definition~4 depending
respectively on whether $X$ is a continuous or a discrete variable, and let us define the sets
$\mathcal{G}[\lambda]$ and $\mathcal{H}[\lambda]$ in the same way but with regard to the variables
$Y$ and $Z$ and the distribution functions $G_Y(y)$ and $H_Z(z)$ respectively.
Now, it would be reasonable to assume in general that, for any three events $A$, $B$ and $C$ that
are members of the sets $\mathcal{F}[\lambda]$, $\mathcal{G}[\lambda]$ and $\mathcal{H}[\lambda]$
respectively, a value of $\alpha \in [0,1]$ could be found that satisfies the condition:
\begin{equation}
\label{equ10}
S(B, R(\lambda)) = S(\, \{\hspace*{0.1em} A \cap L(\alpha)^{c} \hspace*{0.1em}\} \cup
\{\hspace*{0.1em} C \cap L(\alpha) \hspace*{0.1em}\},\, R(\lambda)\,)
\end{equation}
where $L(\alpha)$ is a scaling event that is independent of the variables $X$ and $Z$ and of the
scaling events $L_1(b_1), L_2(b_2), \ldots$ that may have been required to define the sets
$\mathcal{F}[\lambda]$ and $\mathcal{H}[\lambda]$, and where $L(\alpha)^{c}$ denotes the complement
of $L(\alpha)$.
This value of $\alpha$ could then be interpreted as a continuous measure of the external strength
of the distribution function $G_Y(y)$ relative to both the distribution functions $F_X(x)$ and
$H_Z(z)$.

For instance, a small value of $\alpha$ would indicate that the external strength of the function
$G_Y(y)$ is closer to that of the function $F_X(x)$ than to that of the function $H_Z(z)$, while a
large value of $\alpha$ would indicate that the external strength of $G_Y(y)$ is closer to that of
$H_Z(z)$ than to that of $F_X(x)$.
However, it will be assumed that this continuous measure of external strength may only be applied
in cases where it can not or does not contradict the general definition of external strength given
earlier, i.e.\ Definition~7.

\vspace*{2ex}
To give an example of the application of the definition of external strength that has just been
proposed, let us apply this definition to the case considered in Section~\ref{sec7}.
In particular, it would appear acceptable to assume that, in terms of this definition, the function
$F_X(x)$ could be, according to the notation used \pagebreak in this earlier example, the prior or
posterior distribution function $D(\mu)$ or $D(\mu\,|\,x)$, the function $G_Y(y)$ could be the
fiducial distribution function $C(\mu\,|\,x)$, and the function $H_Z(z)$ could be the Bernoulli
distribution function $U_Z(z)$ that corresponds to quite sensibly assigning a probability of
$\omega$ to the event of drawing a ball out of an urn containing $k$ distinctly labelled balls that
belongs to a given subset of $\hspace*{0.1em}\omega k\hspace*{0.1em}$ balls (which will be taken as
being the event that $z=1$) and a probability of $1-\omega$ to the complement of this event (i.e.\
the event that $z=0$).
If it is also assumed that the similarity on the left-hand side of equation~(\ref{equ10}) is
evaluated by using fiducial reasoning, and the similarity on the right-hand side of this equation
is only evaluated directly (and before the data are observed) if $F_X(x)$ is taken to be $D(\mu)$,
or in addition by using Bayesian reasoning if $F_X(x)$ is taken to be $D(\mu\,|\,x)$, then by using
the same type of principles that were explained in Section~\ref{sec7}, it could be argued that
$\alpha$ should be equal or close to one.
If $\alpha=1$, then this would naturally imply that the external strengths of the functions
$C(\mu\,|\,x)$ and $U_Z(z)$ are equal, while if $\alpha$ is close to one, then this could be
interpreted as meaning that, in the context of the reasoning processes being used, the external
strength of $C(\mu\,|\,x)$ is much closer to that of $U_Z(z)$ than to that of the function $D(\mu)$
or $D(\mu\,|\,x)$.

\vspace*{3ex}
\section{Discussion}
\label{sec11}

We will now discuss how well the theory outlined in the present paper addresses the criticisms~1
to~4 listed in the Introduction of the definition of probability that was outlined both in
Bowater~(2017a) under the name `type B probability' and in Bowater~(2017b).

\vspace*{3ex}
\noindent
{\bf Criticism 1: Satisfying the additivity rule of probability}

\vspace*{1ex}
\noindent
Obeying the additivity rule is no longer a goal for the theory, as was the case in Bowater~(2017a),
but rather an assumption upon which the development of a definition of probability is based.
In particular, to guarantee that this assumption is satisfied, the definition of probability put
forward has been centred exclusively on probability distribution functions instead of also centring
it on the probabilities of events in isolation.
Nevertheless, in any given situation, the adequacy of making the assumption that probabilities are
additive is reflected in the relative external strengths that are associated with the distribution
functions concerned.
For example, if in eliciting a distribution function $F_X(x)$ for a given random variable $X$, the
assumption that the probabilities of $X$ lying in distinct (non-overlapping) subsets of the sample
space of $X$ are additive was uncomfortable to make, then we would not expect all the similarities
in the set $\{ S( A, R(\lambda) ) : A \in \mathcal{F}[\lambda] \}$ to be regarded as being close to
the highest similarity that can exist between two events, where the reference set $R$ and the range
of $\lambda$ are again specified as in the examples of Section~\ref{sec5}.

\vspace*{3ex}
\noindent
{\bf Criticism 2: Precision of probability values}

\vspace*{1ex}
\noindent
Unlike in Bowater~(2017a, 2017b), where probabilities were only defined at potentially quite widely
spaced points on the continuous interval $[0,1]$, probabilities can now take any value in this
interval. Observe that this is the case even if the reference set of events $R$ is discrete, as the
precision by which probabilities can be measured is no longer determined by how the set $R$ is
defined, as was the case in these earlier papers.

However, in contrast to this earlier work, there is no guarantee that the probability that is
elicited for any given event is a unique value.
This is because there may be a set $F^{*}$ of possible distribution functions $F_X(x)$ for a given
variable $X$, each member of which is regarded to be internally stronger than any function $F_X(x)$
not in this set, but not internally stronger than any other function $F_X(x)$ within this set.
It would be hoped, though, that usually the distribution functions in the set $F^{*}$ would be
fairly similar to each other. In this type of situation, it is recommendable that any statistical
analysis that requires a distribution function for $X$ as an input incorporates a sensitivity
analysis over the functions $F_X(x)$ in the set $F^{*}$.

\pagebreak
\noindent
{\bf Criticism 3: Dependence of probabilities on the reference set}

\vspace*{1ex}
\noindent
As was the case in Bowater~(2017a, 2017b), probabilities and the strength of probabilities need to
be regarded, in general, as being dependent on the reference set of events $R$ with respect to
which the probabilities concerned are defined.
As far as the present paper is concerned, this means that we need to regard the relative internal
and external strengths of a distribution function as being generally dependent on the set $R$ that
is being used.
As alluded to in Section~\ref{sec8}, this issue though is made substantially less important by
taking into account that reference sets may often be regarded as being compatible for given values
of $\lambda$ according to the definition given in Section~\ref{sec6}.

We could of course attempt to completely remove the dependency in question by defining the
reference set of events $R$ under the added condition that the set of events $O=\{ O_1, O_2,
\ldots, O_k \}$ in the case where $R$ is discrete, or the continuous variable $V$ in the case where
$R$ is continuous, must be the outcomes or outcome of a well-understood physical experiment. For a
continuous reference set $R$, this would mean that the set $R$ would be composed entirely of
scaling events according to the definition in Section~\ref{sec3}. However, placing this extra
condition on the set $R$ would not appear to be that helpful for at least two reasons.

First, since the definition of the set $\mathcal{F}[a]$ for a continuous distribution function
given in equation~(\ref{equ8}) does not depend on the concept of a scaling event, the definitions
of the set $R$ given in Section~\ref{sec1} allow us at least to define the concepts of relative
internal and external strength of a continuous distribution function without entering into a
potentially woolly discussion about when an outcome or set of outcomes can or should be classified
as the outcome or outcomes of a well-understood physical experiment.
The second reason for using these earlier given definitions of the set $R$ is that, in some
situations, it may be useful to base assessments of uncertainty on a set $R$ that does not contain
any event that could be regarded as being an outcome of a simple type of physical experiment.

In particular, if the goal of an individual is to communicate what he knows about a random variable
$X$ to others by not just reporting his elicited distribution function $F_X(x)$ for this variable,
but also explicitly associating his uncertainty about whether events in the set
$\mathcal{F}[\lambda]$ will occur with his uncertainty about whether an event $R(\lambda)$ will
occur, then he may not feel that his choice for the event $R(\lambda)$ is the most appropriate if
the similarities in the set $\{\, S( A, R(\lambda) ) : A \in \mathcal{F}[\lambda] \,\}$ are
generally quite low.
For instance, if this situation arises when the reference set of events $R$ is based on the outcome
of spinning the type of wheel described in Section~\ref{sec1}, then the individual may wish to try
to find an alternative set $R$ that, like the original set $R$, contains events associated with a
standardised form of uncertainty that can be clearly appreciated by many people, but with respect
to which the similarities in the set $\{ S( A, R(\lambda) ) : A \in \mathcal{F}[\lambda] \}$ are
felt by him to be generally higher than they were in the original scenario.

To give an example of when an individual may wish to take this option, let us begin by observing
that if with regard to the governor election example first discussed in Section~\ref{sec5}, the
function $H_Z(z)$ is the already-elicited distribution function over the events $z=\{ z_1, z_2,
\redots, z_5 \}$ of each of the five candidates winning, and if the reference set $R$ is the
original type of set $R$ just mentioned, then as alluded to in Section~\ref{sec4}, it may be
difficult to regard the similarities in the set $\{ S( A, R(\lambda) ) : A \in
\mathcal{H}[\lambda] \}$ as generally being that high, where the set $\mathcal{H}[\lambda]$ is
specified as the set $\mathcal{F}[\lambda]$ was defined in Definition~4 but with respect to the
variable $Z$ and the distribution function $H_Z(z)$.
For this reason, let us now change the reference set of events $R$ in this example to a discrete
reference set $R$ that, similar to a reference set that was mentioned in Section~\ref{sec6}, is
based on the outcomes of drawing a ball out of an urn that contains a given number of balls, each
of which is marked with a number in the range 1 to $k$, but for which the number of balls marked
with any given number is entirely unknown.
Due to the added ambiguity in the uncertainty about whether any given event $R(\lambda)$ in the set
$R$ will occur, it would seem plausible that, with respect to this reference set $R$, a rational
individual could now in fact consider the similarities in the set
$\{ S( A, R(\lambda) ) : A \in \mathcal{H}[\lambda] \}$ as being reasonably high.
Given that also the events in this alternative reference set $R$ are associated with a fairly
standard and easily understood type of uncertainty, an individual may feel it is easier to convey
his opinion about what could be the outcome of the governor election to others by making an analogy
between his uncertainty about whether any given event in the set $\mathcal{H}[\lambda]$ will occur
and his uncertainty about whether the event $R(\lambda)$ in this reference set will occur, rather
than whether the event $R(\lambda)$ in the original reference set will occur.

For a similar reason, if Definition~4 of internal strength is to be used to elicit the distribution
function $H_Z(z)$, then it may well be more appropriate to use this definition under the assumption
that the set $R$ is the alternative reference set in question rather than the original reference
set, and it may perhaps be more convenient to evaluate the relative external strength of the
elicited distribution function with the set $R$ specified in this alternative way.

\vspace*{3ex}
\noindent
{\bf Criticism 4: Lack of universality of the definition}

\vspace*{1ex}
\noindent
The present work has addressed the lack of universality of the definition of probability outlined
in Bowater~(2017a, 2017b) by defining the concepts of internal and external strength so that they
can be applied not just to continuous but also to discrete distribution functions, and in doing
so, thereby eliminating the need for the concept of strength to have a separate definition in the
case where it is applied to the probability of an individual event in isolation, such as the type
of definition that was designated for this specific purpose in these earlier papers.
In particular, with regard to the theory put forward in the present paper, the strength of the
probability $p$ of an individual event can, if desired, be associated with the internal or external
strength of a distribution function that assigns a probability $p$ to this event and a probability
of $1-p$ to the complement of this event, i.e.\ a Bernoulli distribution function.

\vspace*{3ex}
\section{Some closing remarks}

With regard to the overall motivation that was given for the present theory in the Introduction, it
is hoped that this theory, even more than was the case for the earlier theory that was outlined in
Bowater~(2017a, 2017b), gives the concept of probability a natural and useful real-world meaning.
Also, let us recall that it was shown in Section~\ref{sec4} how the present theory can account for
a rational preference for the second urn in Ellsberg's two urn example, and that furthermore it was
shown in Sections~\ref{sec7} and~\ref{sec9} how, by accounting for the same general issue that
underlies such a preference, it can justify the use of fiducial rather than Bayesian reasoning in
the examples that were discussed in these earlier two sections.
Therefore, it is hoped that this theory has adequately achieved all the goals that were set out at
the start of this paper.

\vspace*{5ex}
\noindent
{\bf References}

\begin{description}

\setlength{\itemsep}{0.75ex}

\item[] Becker, S. W. and Brownson, F. O. (1964).\ What price ambiguity?\ or the role of ambiguity
in decision-making.\ \emph{Journal of Political Economy}, {\bf 72}, 62--73.

\item[] Bowater, R. J. (2017a).\ A formulation of the concept of probability based on the use of
experimental devices.\ \emph{Communications in Statistics:\ Theory and Methods}, {\bf 46},
4774--4790.

\item[] Bowater, R. J. (2017b).\ A defence of subjective fiducial inference.\ \emph{AStA Advances
in Statistical Analysis}, {\bf 101}, 177--197.

\item[] Bowater, R. J. (2018).\ Multivariate subjective fiducial inference.\ \emph{arXiv.org
(Cornell University), Statistics}, arXiv:1804.09804.

\item[] Curley, S. P. and Yates, J. F. (1989).\ An empirical evaluation of descriptive models of
ambiguity reactions in choice situations.\ \emph{Journal of Mathematical Psychology}, {\bf 33},
397--427.

\item[] Eagle, A. (2011).\ \emph{Philosophy of Probability:\ Contemporary Readings}, Routledge,\\
London.

\item[] Ellsberg, D. (1961).\ Risk, ambiguity and the Savage axioms.\ \emph{Quarterly Journal of
Economics}, {\bf 75}, 643--669.

\item[] Fellner, W. (1961).\ Distortion of subjective probabilities as a reaction to uncertainty.\
\emph{Quarterly Journal of Economics}, {\bf 75}, 670--689.

\item[] Fine, T. L. (1973).\ \emph{Theories of Probability:\ An Examination of Foundations},
Academic Press, New York.

\item[] Garthwaite, P. H., Kadane, J. B. and O'Hagan, A. (2005).\ Statistical methods for eliciting
probability distributions.\ \emph{Journal of the American Statistical Association}, {\bf 100},
680--700.

\item[] Gillies, D. (2000).\ \emph{Philosophical Theories of Probability}, Routledge, London.

\item[] Kadane, J. B. and Wolfson, L. J. (1998).\ Experiences in elicitation.\ \emph{Journal of the
Royal Statistical Society, Series D}, {\bf 47}, 3--19.

\item[] Kynn, M. (2008).\ The `heuristics and biases' bias in expert elicitation.\ \emph{Journal of
the Royal Statistical Society, Series A}, {\bf 171}, 239--264.

\item[] O'Hagan, A., Buck, C. E., Daneshkhak, A., Eiser, J. R., Garthwaite, P. H., et al.\ (2006).\
\emph{Uncertain Judgements:\ Eliciting Experts' Probabilities}, Wiley, Chichester.

\item[] Spetzler, C. S. and Stael von Holstein, C. A. S. (1975).\ Probability encoding in decision
analysis.\ \emph{Management Science}, {\bf 22}, 340--358.

\end{description}

\end{document}